\documentclass[twocolumn,aps,prd,showpacs,10pt,superscriptaddress]{revtex4-1}
\usepackage{amssymb}
\usepackage{amsmath}
\newcommand{\vect}[1]{\boldsymbol{#1}} 
\usepackage{multirow}
\usepackage{graphicx}
\usepackage[usenames,dvipsnames,svgnames]{xcolor}
\usepackage{hyperref}
\hypersetup{
pdfnewwindow=true,      
colorlinks=true,        
linkcolor=Blue,         
citecolor=Blue,         
filecolor=Blue,         
urlcolor=Blue           
}

\begin{document}

\title{A minimal Tersoff potential for diamond silicon with improved descriptions of elastic and phonon transport properties}

\author{Zheyong Fan}
\email{brucenju@gmail.com}
\affiliation{School of Mathematics and Physics, Bohai University, Jinzhou, P. R. China}
\affiliation{QTF Centre of Excellence, Department of Applied Physics, Aalto University, FI-00076 Aalto, Finland}
\author{Yanzhou Wang}
\affiliation{Beijing Advanced Innovation Center for Materials Genome Engineering, University of Science and Technology Beijing, Beijing, 100083, China}
\affiliation{School of Mathematics and Physics, University of Science and Technology Beijing, Beijing 100083, P. R. China}
\author{Xiaokun Gu}
\email{Xiaokun.Gu@sjtu.edu.cn}
\affiliation{Institute of Engineering Thermophysics, School of Mechanical Engineering, Shanghai Jiao
Tong University, Shanghai 200240, China}
\author{Ping Qian}
\affiliation{Beijing Advanced Innovation Center for Materials Genome Engineering, University of Science and Technology Beijing, Beijing, 100083, China}
\affiliation{School of Mathematics and Physics, University of Science and Technology Beijing, Beijing 100083, P. R. China}
\author{Yanjing Su}
\email{yjsu@ustb.edu.cn}
\affiliation{Beijing Advanced Innovation Center for Materials Genome Engineering, University of Science and Technology Beijing, Beijing, 100083, China}
\author{Tapio Ala-Nissila}
\affiliation{QTF Centre of Excellence, Department of Applied Physics, Aalto University, FI-00076 Aalto, Espoo, Finland}
\affiliation{Centre for Interdisciplinary Mathematical Modeling and Department of Mathematical Sciences, Loughborough University, Loughborough, Leicestershire LE11 3TU, UK}

\date{\today}

\begin{abstract}
Silicon is an important material and many empirical interatomic potentials have been developed for atomistic simulations of it. Among them, the Tersoff potential and its variants are the most popular ones. However, all the existing Tersoff-like potentials fail to reproduce the experimentally measured thermal conductivity of diamond silicon. Here we propose a modified Tersoff potential and develop an efficient open source code called GPUGA (graphics processing units genetic algorithm) based on the genetic algorithm and use it to fit the potential parameters against energy, virial and force data from quantum density functional theory calculations. This potential, which is implemented in the efficient open source GPUMD (graphics processing units molecular dynamics) code, gives significantly improved descriptions of the thermal conductivity and phonon dispersion of diamond silicon as compared to previous Tersoff potentials and at the same time well reproduces the elastic constants. Furthermore, we find that quantum effects on the thermal conductivity of diamond silicon at room temperature are non-negligible but small: using classical statistics underestimates the thermal conductivity by about 10\% as compared to using quantum statistics.
\end{abstract}

\maketitle

\section{Introduction}

Thermal transport in silicon based materials has been extensively studied by classical molecular dynamics (MD) simulations \cite{volz2000prb,henry2008jctn,donaido2009prl,donaido2010nl,lampin2012apl,howell2012jcp,xiong2014prb,saaskilahti2016aipa,cartoix2016apl,zaoui2017prb,zhou2017nl,dong2018prb}. The results strongly depend on the empirical interatomic potential used. Quantitatively accurate empirical potentials for covalently bonded solids such as silicon are many-body in nature and cannot be expressed as sums of pairwise interactions. Among the various many-body empirical potentials, the Tersoff potential \cite{tersoff1988prb,tersoff1989prb} is the most frequently used for silicon. In addition to the original parametrizations by Tersoff \cite{tersoff1988prb,tersoff1989prb}, this potential has also been modified and/or re-parametrized by many other authors \cite{erhart2005prb,kumagai2007cms,Pun2017prb}. Although the Tersoff potential has a relatively simple form and low computational cost compared to many other many-body potentials, it can capture the essence of quantum-mechanical bonding \cite{brener2005}, justifying its widespread use in modeling \cite{mota1998prb,albe2002prb_1,albe2002prb,nord2003jpcm,erhart2006jpcm,munetoh2007cms,muller2007jpcm,powell2007prb,Henriksson2009prb,los2017prb,Byggmastar2018jpcm}.

\begin{figure}[htb]
\begin{center}
\includegraphics[width=\columnwidth]{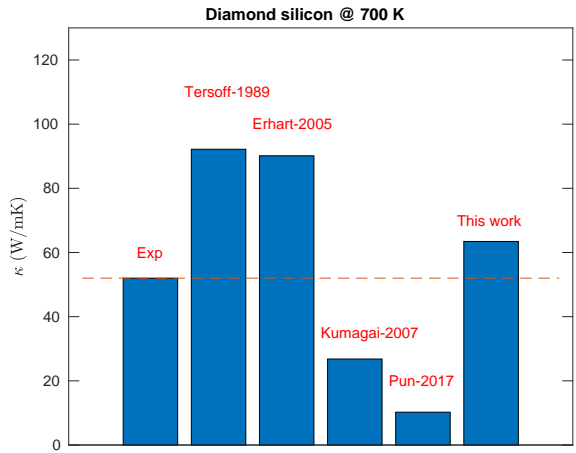}
\caption{Thermal conductivity of diamond silicon at 700 K and zero pressure from experiments \cite{Glassbrenner1964pr} and some commonly used empirical potentials using the homogeneous nonequilibrium molecular dynamics method \cite{Fan2019prb} as implemented in the GPUMD code \cite{fan2017cpc,gpumd}. Isotope scattering is taken into account here. See text for details.}
\label{figure:sw_tersoff}
\end{center}
\end{figure}

There are however some features that cannot be consistently reproduced by Tersoff-like potentials, such as heat conductivity in the solid phase. In particular, in Fig. \ref{figure:sw_tersoff} we show the thermal conductivity of the standard diamond silicon structure at 700 K (where quantum effects can be neglected) and zero pressure predicted by using the previous Tersoff-like potentials as well as the one introduced in this work. None of the previous ones gives a reasonable match to the reference experimental value of $51$ W/mK. The Tersoff potentials parametrized by Tersoff \cite{tersoff1989prb} and Erhart and Albe \cite{erhart2005prb} (the one named as Si-II which was suggested to be better for simulation with elemental silicon) predict comparable values and overshoot the experimental value by about $80\%$, while the modified Tersoff potentials by Kumagai \textit{et al.} \cite{kumagai2007cms} (the one called MOD in this reference) and by Pun and Mishin \cite{Pun2017prb} underestimate the experimental value by a factor of 2 and 5, respectively. The fact that the Tersoff-like potentials can predict very different thermal conductivity values suggests that accurate prediction of the thermal conductivity could be achieved with an appropriate functional form and parametrization.

To achieve this goal, we propose here a minimal Tersoff potential for silicon. Here, by ``minimal'' we mean that every parameter in the potential is essential and there is no redundancy. In the original Tersoff potential \cite{tersoff1989prb}, there are 11 parameters. The version used by Erhart and Albe \cite{erhart2005prb} has the same number of parameters, although some functions have been written in a different but equivalent way. In the modified Tersoff potential by Kumagai \textit{et al.} \cite{kumagai2007cms}, 16 parameters were used, and the latest version by Pun and Mishin \cite{Pun2017prb} used 17. Here, instead of going with this trend of increasing the number of fitting parameters and the complexity of the potential, we do the opposite. With extensive fitting trials with the help of a genetic algorithm, we find that three parameters in the Tersoff potential can be eliminated without adversely affecting the quality of the fitting. A set of optimized parameters were found by fitting the minimal Tersoff potential against energy, virial, and force data from quantum density functional theory (DFT) \cite{hohenberg1964pr,kohn1965pr} calculations for many configurations. Our optimized potential predicts a thermal conductivity which only overshoots the experimental value by about $20\%$ at 700 K.

This paper is organized as follows. In Sec. \ref{section:potential}, we introduce the functional form of the minimal Tersoff potential. In Sec. \ref{section:fitting}, we present the details of the training data and the fitting method. In Sec. \ref{section:optimized}, we evaluate the optimized potential in terms of elastic constants, phonon dispersion, and thermal conductivity. In Sec. \ref{section:summary} we present our summary and conclusions.

\section{Potential model\label{section:potential}}

\subsection{The Tersoff potential}

We first briefly introduce the Tersoff potential in the form published by Tesoff in 1989 \cite{tersoff1989prb}. This is equivalent to the form used by Erhart and Albe in 2005 \cite{erhart2005prb}.

The total potential energy (cohesive energy) $U$ for a system with $N$ atoms is written as a sum the site potentials:
\begin{equation}
U = \sum_{i=1}^N U_{i}.
\end{equation}
The site potential for atom $i$ is formally written as
\begin{equation}
U_{i}=\frac{1}{2} \sum_{j\neq i}^N U_{ij},
\end{equation}
where the potential between atoms $i$ and $j$ is
\begin{equation}
U_{ij}= f_{\rm C}(r_{ij}) \left[ f_{\rm R}(r_{ij}) - b_{ij} f_{\rm A}(r_{ij}) \right].
\end{equation}
This is the general form of the Tersoff potential. Here, $f_{\rm C}(r_{ij})$ is the pairwise cutoff function, $f_{\rm R}(r_{ij})$ and $f_{\rm A}(r_{ij})$ are respectively the pairwise repulsive and attractive functions, and $b_{ij}$ (not equal to $b_{ji}$ in general) is the bond order for the $ij$ bond. Many-body effects are totally embodied in the bond order. The repulsive and attractive functions take the following forms:
\begin{equation}
f_{\rm R}(r_{ij}) = A e^{-\lambda r_{ij}};
\label{equation:f_R}
\end{equation}
\begin{equation}
f_{\rm A}(r_{ij}) = B e^{-\mu r_{ij}},
\label{equation:f_A}
\end{equation}
where $A>0$, $\lambda>0$, $B>0$, $\mu>0$ are fitting parameters.

The bond order $b_{ij}$ is expressed as
\begin{equation}
b_{ij}=\left(1+\zeta_{ij}^n\right)^{-1/2n};
\end{equation}
\begin{equation}
\zeta_{ij}=\sum_{k\neq i,j}^N f_C({r_{ik}}) g(\theta_{ijk}),
\end{equation}
where $n>0$ is a fitting parameter. A larger $\zeta_{ij}$ gives a smaller $b_{ij}$ and a weaker bond. When $\zeta_{ij}=0$, $b_{ij}$ attains a maximum value of one. The angular function is chosen as
\begin{equation}
g^{\rm T}(\theta_{ijk}) = \gamma
\left(
1 + \frac{c^2}{d^2} - \frac{c^2}{d^2 + \left( \cos\theta_{ijk} - h \right)^2}
\right),
\label{equation:g_tersoff}
\end{equation}
where $\gamma$, $c$, $d$ and $h$ are fitting parameters and $\theta_{ijk}$ is the bond angle formed by the $ij$ and $ik$ bonds.

In the expressions of $U_{ij}$ and $\zeta_{ij}$, there is a cutoff function $f_{\rm C}(r)$ which takes the following form:
\begin{equation}
f_{\rm C}(r) =
\begin{cases}
    1, & r\leq R_1; \\
    \frac{1}{2}\left[ 1+\cos\left(\pi\frac{r-R_1}{R_2-R_1}\right)\right], & R_1<r<R_2; \\
    0, & r\geq R_2.
\end{cases}
\label{equation:fc_tersoff}
\end{equation}
Here, $R_1>0$ and $R_2>R_1$ are the inner and outer cutoff distances, respectively.

The cutoff distances $R_1$ and $R_2$ are usually not optimized systematically but are chosen by hand instead. Therefore, there are 9 fitting parameters for the Tersoff potential: $A$, $B$, $\lambda$, $\mu$, $\gamma$, $n$, $c$, $d$, $h$.

\subsection{The minimal Tersoff potential}

We note that in most Tersoff potentials, $c^2/d^2\gg 1$, $d^2\gg 1$, and $\gamma \ll 1$. Under these conditions, $g^{\rm T}(\theta_{ijk}) \approx \gamma c^2/d^4 \left( \cos\theta_{ijk} - h \right)^2$. Defining $\beta=\gamma c^2/d^4$, we obtain
\begin{equation}
g(\theta_{ijk}) = \beta \left( \cos\theta_{ijk} - h \right)^2.
\label{equation:g}
\end{equation}
An advantage of Eq. (\ref{equation:g}) over Eq. (\ref{equation:g_tersoff}) is that the fitting parameter $\beta$ takes a value of the order of unity, while those in the original Tersoff potential take values differing by orders of magnitude. Therefore, our new angular function is much easier to fit.

As in most previous Tersoff potentials, we do not fit $R_1$ and $R_2$ but chose their values by hand. We choose $R_1=2.8$ \AA ~and $R_2=3.2$ \AA, but they can be modified as needed. None of the training data involve atom pairs with distances within the two cutoffs. One of the drawbacks of the bond-order potentials is the abrupt cutoff function, which results in abnormally large forces when two atoms are within the two cutoff distances. Screened bond-order potentials \cite{Pastewka2008prb,Pastewka2013prb,perriot2013prb} have been proposed overcome this drawback. Because our focus here is on the elastic and thermal properties of diamond silicon, we do not consider these advanced cutoff schemes. 

In our numerical implementation, we do not fit the parameters $A$, $B$, $\lambda$, and $\mu$ directly, but instead translate them to another set of equivalent parameters $D_0$, $\alpha$, $r_0$, and $S$ as done by Erhart and Albe \cite{erhart2005prb}:
\begin{equation}
    A=\frac{D_0}{S-1} \exp\left(\alpha r_0\sqrt{2S} \right);
\end{equation}
\begin{equation}
    B=\frac{D_0S}{S-1} \exp\left(\alpha r_0\sqrt{2/S} \right);
\end{equation}
\begin{equation}
    \lambda= \alpha\sqrt{2S};
\end{equation}
\begin{equation}
    \mu= \alpha\sqrt{2/S}.
\end{equation}
The advantage of using the parameters $D_0$, $\alpha$, $r_0$, and $S$ in the fitting process is that they all have values of the order of unity (when energy is in units of eV and length is in units of \AA), which makes it easier to set up ranges for their allowed values. Physically, $S$ is the slope parameter in the Pauling plot (bond energy versus bond length). When $S=2$, the combination of the repulsive and attractive functions in Eqs. (\ref{equation:f_R}) and (\ref{equation:f_A}) reduces to the Morse function. In our fitting trials, we always got $S \approx 2$ (up to $0.1\%$ deviation only) and we thus fix $S= 2$ and do not treat it as a fitting parameter. Therefore, there are only 6 fitting parameters for our minimal Tersoff potential: $D_0$, $\alpha$, $r_0$, $\beta$, $n$, $h$. To our knowledge this is a Tersoff-like potential with the smallest number of fitting parameters proposed so far.

\section{Fitting database and fitting method  \label{section:fitting}}

\subsection{DFT calculations for the training data}

DFT calculations are performed using the Vienna Ab initio Simulation Package (VASP) \cite{Kresse1996prb} that employs a plane-wave basis (we chose a kinetic energy cutoff of 600 eV) and the projector augmented wave (PAW) method \cite{blochl1994prb,kresse1999prb}. A $\Gamma$-centered uniform $k$-point grid with $16 \times 16 \times 16$ $k$-points is employed in the total energy calculations for a cubic unit cell with 8 atoms and a similar $k$-point density is used for other unit cells. When atom positions are to be optimized, the stopping criteria is to make the force on each atom smaller than $10^{-4}$ eV/\AA. Spin is considered in all the calculations. As for the exchange-correlation energy functional, we have compared the following three variants: local spin density approximation (LSDA) \cite{perdew1992prb2},  generalized-gradient approximation \cite{perdew1992prb} as parametrized by Perdew, Burke, and Ernzerhof (GGA-PBE) \cite{perdew1996prl}, and a revised GGA-PBE (GGA-PBEsol) \cite{perdew2008prl}. We first performed a calculation with the atom positions allowed to be optimized. The calculated lattice constant and the corresponding cohesive energy for the diamond structure using different functionals are listed in Table \ref{table:lattice_and_cohesive}. While GGA-PBE gives the most accurate cohesive energy compared with the experimental data, GGA-PBEsol gives the most accurate lattice constant. As we will shift the energy before fitting the potential parameters (see below), we chose to use the results from GGA-PBEsol in the fitting.

\begin{table}[htb]
\centering
\caption{Lattice constants ($a$) and cohesive energies ($E_{\rm c}$) for diamond silicon calculated by using different exchange-correlation functionals.  }
\begin{tabular}{lll}
\hline
Functional & $a$ (\AA)   & $E_{\rm c}$ (eV/atom)   \\
\hline
\hline
LSDA      & $5.403$      & $-5.35$  \\
\hline
GGA-PBE   & $5.469$      & $-4.61$  \\
\hline
GGA-PBEsol & $5.436$     & $-4.93$   \\
\hline
Experimental & $5.43$  & $-4.63$  \\
\hline
\hline
\end{tabular}
\label{table:lattice_and_cohesive}
\end{table}

After obtaining the ground state structure, we create unit cells with  triaxial,  biaxial,  and uniaxial  deformations.  For  each  deformation  type,  we  consider  strains $\epsilon$ from $-10\%$ to $10\%$, with smaller steps around the ground state. For each structure, we calculate the total energy and virial tensor without optimizing the atom positions (single-point calculations). The calculated cohesive energy (from GGA-PBEsol) for the ground state deviates from the experimental value to some degree. This might be related to the difficulty of accurately determining the energy of an isolated atom. In order to obtain an empirical potential that can reproduce the experimental value of the cohesive energy, we shift all the DFT energies by a constant value such that the ground state cohesive energy is $-4.63$ eV per atom. Similar corrections were made by Kumagai \textit{et al.} \cite{kumagai2007cms}.

To increase the the diversity of bond angles and coordination numbers in the training database, we also consider a few (artificial or real) allotropes of silicon: simple cubic crystal, body-centered cubic crystal, face-centered cubic crystal, and two-dimensional silicene at their ground states with zero stress. Apart from energy and virial, we also include the forces in the training database. To this end, we use the Tersoff potential \cite{tersoff1989prb} to generate five configurations at 100, 200, 300, 400, and 500 K, and calculate the force on each atom using DFT. The system here is a cubic cell consisting of 64 atoms with periodic boundary conditions in all directions.

\subsection{Genetic algorithm as the fitting method}

Simultaneously optimizing all the parameters is a challenging task for conventional fitting methods, but a metaheuristic such as the genetic algorithm (GA) is well suited to handle it. The GA is a global optimization method and has been successfully used in some previous works to optimize complex potentials with many parameters \cite{larsson2013jcc,kumagai2007cms,rohskopf2017npj}. Other global optimization methods such as the particle swarm optimization method has also been used to fit empirical potentials \cite{kandemir2016nt}. Here, we use the GA to optimize all the parameters in our potential simultaneously. 

In our optimization problem, the fitness function (also called objective or cost function) to be minimized is a weighted sum of the errors for energy, virial and force:
\begin{equation}
    Z(x) = w_{\rm e} Z_{\rm e}(x) + w_{\rm v} Z_{\rm v}(x) + w_{\rm f} Z_{\rm f}(x).
\end{equation}
Here $x$ represents a solution of the optimization problem, which is an array consisting of the potential parameters:
\begin{equation}
    x=[D_0, \alpha, r_0, \beta, n, h].
\end{equation}
For energy, we define the fitness function as
\begin{equation}
    Z_{\rm e}(x) = \left(\frac{\sum_n|E(n) - E^{\rm DFT}(n) |^2}{\sum_n |E^{\rm DFT}(n)|^2} \right)^{1/2},
\end{equation}
where $E(n)$ and $E^{\rm DFT}(n)$ are the energies of the $n$-th structure calculated from the empirical potential and DFT, respectively. Similarly, the fitness function for virial is defined as
\begin{equation}
    Z_{\rm v}(x) = \left(\frac{\sum_n\sum_{\mu\nu}|\sigma_{\mu\nu}(n) - \sigma_{\mu\nu}^{\rm DFT}(n) |^2}{\sum_n \sum_{\mu\nu} |\sigma^{\rm DFT}_{\mu\nu}(n)|^2} \right)^{1/2},
\end{equation}
where $\sigma_{\mu\nu}(n)$ and $\sigma^{\rm DFT}_{\mu\nu}(n)$ are the $\mu\nu$ virial component of the $n$-th structure calculated from the empirical potential and DFT, respectively. The $\mu\nu$ summation is over the nonequivalent elements of the second-rank virial tensor. The fitness function for force is
\begin{equation}
    Z_{\rm f}(x) = \left(\frac{\sum_n\sum_{i}|\vect{f}_{i}(n) - \vect{f}^{\rm DFT}_i(n) |^2}{\sum_n \sum_{i} |\vect{f}^{\rm DFT}_{i}(n)|^2} \right)^{1/2},
\end{equation}
where $\vect{f}_{i}(n)$ and $\vect{f}^{\rm DFT}_{i}(n)$ are the force on the $i$-th atom in the $n$-th structure calculated from the empirical potential and DFT, respectively. The above fitness function is similar to those used in the potfit \cite{brommer2015msmse} and POPS \cite{rohskopf2017npj} packages. 

The weighting factors $w_{\rm e}$, $w_{\rm v}$ and $w_{\rm f}$ can be adjusted to control the relative emphasis on the targeting properties.  A smaller $Z$ corresponds to a better solution. This unambiguous criteria is the basis for applying the GA.

The workflow of the GA we used is as follows:
\begin{enumerate}
    \item Initialization. Create $N_{\rm pop}$ individual solutions $\{x_i\}_{i=1}^{N_{\rm pop}}$, which form a population with population size $N_{\rm pop}$. In this work, we use a real-valued chromosome representation, where each gene in a chromosome represents a potential parameter. Therefore, there are $N_{\rm pop}$ chromosomes in each generation, and each chromosome has $6$ genes. The translation between the genotype and the phenotype is very simple: each gene takes a value within $[0, 1]$, which is translated to a potential parameter according to two limiting values we set for that parameter.
    \item Loop over $N_{\rm gen}$ generations
    \begin{enumerate}
        \item Evaluate the fitness functions $Z(x_i)$ for all the individuals $x_i$ in the population, sorting them according to the fitness values.
        \item Keep the best solution (the elite) in each generation without altering it.
        \item Select $N_{\rm par}$ individuals with better fitness (smaller $Z$ values) as parents and discarding the remaining ones.
        \item Perform the crossover genetic operation on the $N_{\rm par}$ selected parents, producing $N_{\rm pop}-N_{\rm par}$ new individuals (children) such that the population size is recovered.
        \item Randomly choose some genes in some chromosomes with a given probability and mutate them, i.e., change their values randomly.
    \end{enumerate}
\end{enumerate}

After trial and error, we found that the following parameters are good choices: $N_{\rm pop}=200$, $N_{\rm par}=100$, $N_{\rm gen}=1000$, and a mutation rate linearly decreasing from $0.2$ to zero during the genetic evolution.

\subsection{GPU implementation}

While the GA is generally capable of finding globally optimized potential parameters, it requires evaluating the fitness function many times. It is therefore desirable to make an efficient computer implementation.

Recently, efficient implementation of force evaluation routines in graphics processing units (GPU) has been made for general many-body potentials \cite{fan2017cpc,gpumd}. However, it has also been demonstrated that the computational speed sensitively depends on the simulation cell size. In the calculations here, we only need to use a small simulation cell containing $N_{\rm a}=64$ silicon atoms to incorporate all the interactions. For a system as small as this, a naive GPU implementation barely results in a speedup compared to a CPU implementation. To overcome this difficulty, we note that in each generation, we have $N_{\rm pop}$ individuals, each corresponding to $N_{\rm c}$ configurations. We thus have $N_{\rm pop}N_{\rm c}$ configurations in each generation, which are independent of each other. Therefore, we can use a single CUDA kernel to calculate the physical properties (energy, force, and virial) of part or all of the configurations. The effective system size for the CUDA kernel is thus large enough to achieve a considerable speedup. With our efficient GPU code, performing one optimization with $1000$ generations only takes a few minutes using a Tesla P100 graphics card. This allows us to do a huge number of fitting trials. The fitting code is called GPUGA and it is publicly available \cite{gpuga}.

\begin{table}
\centering
\caption{Optimized parameters of the minimal Tersoff potential for silicon systems.}
\begin{tabular}{llrrr}
\hline
Parameter & Units   &  Value\\
\hline
\hline
$D_0$     & eV         & $3.21481$\\
\hline
$\alpha$       & \AA$^{-1}$ & $1.43134$\\
\hline
$r_0$     & \AA        & $2.23801$\\
\hline
$\beta$     & Dimensionless        & $0.282818$ \\
\hline
$n$       & Dimensionless          & $0.602568$\\
\hline
$h$       & Dimensionless          & $-0.641048$\\
\hline
$R_1$     & \AA        & $2.8$ \\
\hline
$R_2$     & \AA        & $3.2$ \\
\hline
\hline
\end{tabular}
\label{table:optimized_parameters}
\end{table}

\begin{figure*}[htb]
\begin{center}
\includegraphics[width=1.8\columnwidth]{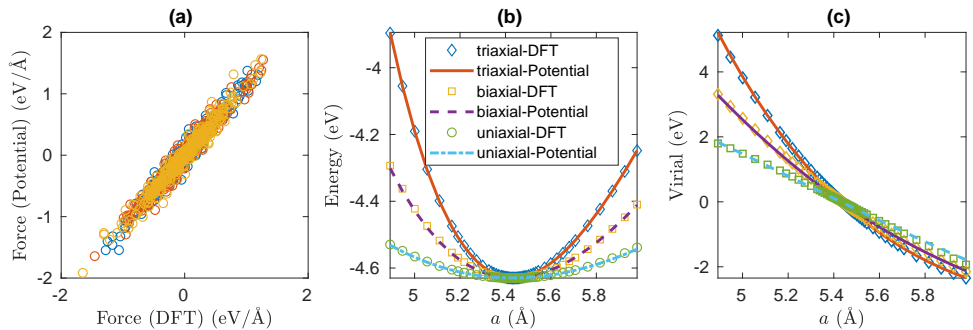}
\caption{(a) Force, (b) energy, and (c) virial as calculated from the minimal Tersoff potential compared with the training data from DFT. In the legend, ``triaxial'' means deforming the three axes of a cubic unit cell by the same amount, ``biaxial'' means deforming two axes only and ``uniaxial'' means deforming one axis only. The lattice constant $a$ in (b) and (c) refers to the deformed value.}
\label{figure:fitting}
\end{center}
\end{figure*}

\subsection{The optimized minimal Tersoff potential }

The optimized parameters for the minimal Tersoff potential are listed in Table \ref{table:optimized_parameters}. Energy, virial, and force calculated using the optimized potential are compared with the DFT training data in Fig. \ref{figure:fitting}. The force and virial stress from the empirical potential were calculated using the formulas in Ref. \cite{fan2015prb}. The agreement with DFT results is reasonably good. The errors for energy and virial are of the order of $1\%$. The cohesive energy and lattice constant calculated using the optimized potential are $-4.63$ eV per atom and $5.434$ \AA, respectively. The error for force is relatively large. It is possible to reduce this error, but at the expense of increasing the errors for energy and virial, resulting in unreasonable elastic constants.

\begin{figure}[htb]
\begin{center}
\includegraphics[width=\columnwidth]{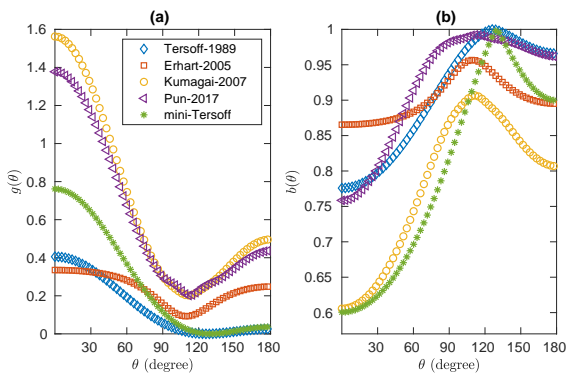}
\caption{The angular function $g$ and bond order $b_{ij}$ as a function of the bond angle $\theta_{ijk}$ in a single triplet $ijk$ for the Tersoff-like potentials considered in this work (see text for details).}
\label{figure:g}
\end{center}
\end{figure}

To see how the current potential differs from previous Tersoff-like potentials, we plot the angular function $g(\theta)$ and the bond order function $b(\theta)$ for a single triplet in Fig. \ref{figure:g}. Our angular function resembles the spline function constructed by Schall \textit{et al.} based on energies in structures with some special bond angles \cite{Schall2008prb}.

\section{Evaluation of the optimized minimal Tersoff potential \label{section:optimized}}

In this section, we evaluate the optimized minimal Tersoff potential in terms of mechanical and thermal properties. We implement this potential into the efficient open-source GPUMD package \cite{fan2017cpc,gpumd} and use this package to do all the MD simulations. We will compare the results with some of the existing Tersoff-type potentials \cite{tersoff1989prb,erhart2005prb,kumagai2007cms,Pun2017prb}.

\subsection{Elastic constants}

\begin{table}
\centering
\caption{Elastic constants (in units of GPa) of diamond silicon from experiments, DFT calculations, and various empirical potentials.
}
\begin{tabular}{lrrrrr}
\hline
Method/Potential  & Taken from & $C_{11}$&  $C_{12}$  & $C_{44}$  & $C_{12}-C_{44}$ \\
\hline
\hline
Experimental \cite{mcskimin1951pr}  & \cite{mcskimin1951pr} & $167.4$  & $65.2$ & $79.6$ &   $-14.4$ \\
\hline
SW  \cite{stillinger1985prb}  & \cite{Pun2017prb} & $151.4$  & $76.4$ & $56.4$ & $20$\\
\hline
Tersoff  \cite{tersoff1988prb}  & \cite{kumagai2007cms} & $142.5$  & $75.4$ & $69.0$ & $6.4$\\
\hline
Erhart (Si-II) \cite{erhart2005prb}  & \cite{erhart2005prb} & $167$  & $65$ & $72$ & $-7$ \\
\hline
Kumagai \cite{kumagai2007cms} & \cite{kumagai2007cms}  & $166.4$  & $65.3$ & $77.1$ & $-11.8$ \\
\hline
Pun \cite{Pun2017prb} & \cite{Pun2017prb} & $172.6$  & $64.6$ & $81.3$ & $-16.7$ \\
\hline
mini-Tersoff  & here & $148$  & $65$ & $75$ &  $-10$ \\
\hline
DFT  & here & $156$  & $62$ & $74$ & $-12$ \\
\hline
\hline
\end{tabular}
\label{table:elastic}
\end{table}

The elastic constants calculated using stress-strain relations at zero temperature are presented in Table \ref{table:elastic}. Our minimal Tersoff potential can predict the correct sign of $C_{12}-C_{44}$, while the SW potential \cite{stillinger1985prb} and the Tersoff-1988 potential \cite{tersoff1988prb} fail. The other potentials \cite{erhart2005prb,kumagai2007cms,Pun2017prb} all describe the elastic properties very well. From Table \ref{table:elastic} and Fig. \ref{figure:sw_tersoff}, we see that there is no clear correlation between the elastic constants and the thermal conductivity. The good elastic properties of our minimal Tersoff potential is implied by the good fit to the energy and virial data in many deformed structures, as shown in Fig. \ref{figure:fitting}.

\subsection{Phonon dispersion}

\begin{figure*}[htb]
\begin{center}
\includegraphics[width=2\columnwidth]{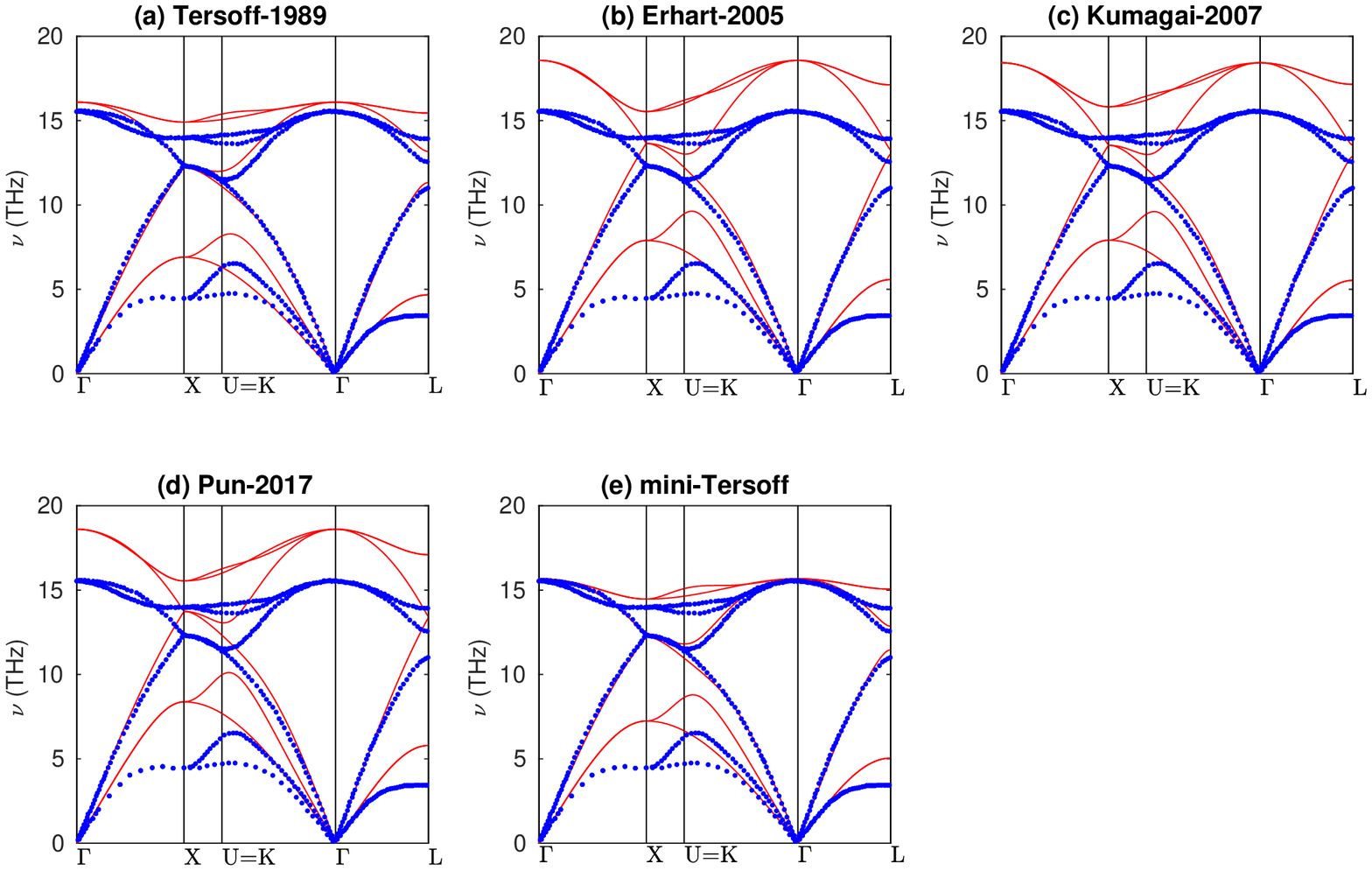}
\caption{Phonon dispersion of diamond silicon from the various empirical potentials (red solid lines) compared with experimental data (blue dots) \cite{holt1999prl}.  }
\label{figure:phonon}
\end{center}
\end{figure*}

To properly describe the phonon transport properties, an adequate description of the phonon dispersion curves is needed. Figure \ref{figure:phonon} shows the phonon dispersions calculated using harmonic lattice dynamics with the second order force constants being calculated from the various empirical potentials using the finite displacement method, compared with experimental data  \cite{holt1999prl} from X-Ray transmission scattering. Here the \verb"phonon" executable within the GPUMD package \cite{gpumd} is used. All of the empirical potentials give a reasonable description for the acoustic branches. However, except for the Tersoff-1989 potential \cite{tersoff1989prb} and our minimal Tersoff potential, all the other potentials give rise to too large a cutoff frequency for the optical branches. Overall, our minimal Tersoff potential gives the best description for the phonon dispersion of diamond silicon among all the empirical potentials considered here.

\subsection{Thermal conductivity}

We next calculate the thermal conductivity $\kappa$ using the efficient homogeneous nonequilibrium molecular dynamics (HNEMD) method \cite{evans1982pla} for many-body potentials \cite{Fan2019prb}. In this method, one generates a non-equilibrium heat current by adding a small external driving force and measure the heat current, which is directly proportional to the thermal conductivity. For details on the HNEMD method, see Ref. \cite{Fan2019prb}.

\begin{figure}[htb]
\begin{center}
\includegraphics[width=\columnwidth]{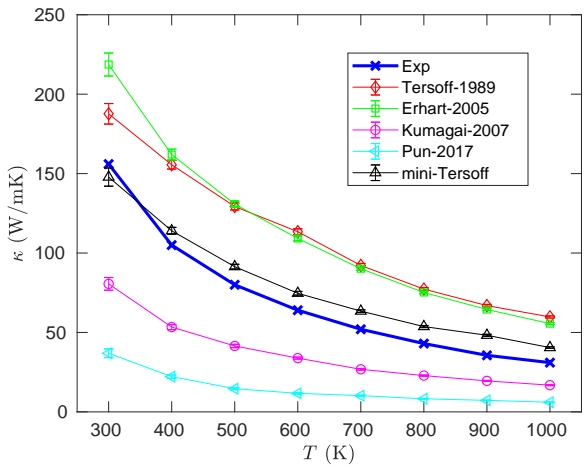}
\caption{Thermal conductivity of diamond silicon as a function of temperature from the various potentials. Isotope scattering is included here. Experimental data are from Ref. \cite{Glassbrenner1964pr}.}
\label{figure:kappa_T}
\end{center}
\end{figure}

We use a simulation cell with 8000 silicon atoms (with periodic boundaries in all three directions) and consider temperatures from 300 to 1000 K, all with zero pressure. To be consistent with experiments, isotope scattering is considered by randomly choosing the mass of a silicon atom according to the following abundance distribution: $92.2\%$ $^{28}$Si, $4.7\%$ $^{29}$Si, and $3.1\%$ $^{30}$Si.

The results obtained by the various Tersoff-like potentials are shown in Fig. \ref{figure:kappa_T} and are compared with experimental data \cite{Glassbrenner1964pr}. Results for $T=700$ K have also been shown in Fig. \ref{figure:sw_tersoff}. It is clear that our minimal Tersoff potential gives results closest to the experimental data. At high temperatures where quantum effects are not important, our predictions are only about $20\%$ \emph{larger} than the experimental values. However, our predicted thermal conductivity at $T=300$ K is slightly \emph{smaller} than the experimental value. This indicates the presence of quantum effects at low temperatures, as we will discuss below.

To explore the influence of quantum effects, we calculate the thermal conductivity by iteratively solving the Peierls-Boltzmann transport equation (PBTE). In our calculations, we consider both three-phonon and four-phonon scatterings \cite{gu2019prb} and temperature-dependent interatomic force constants \cite{hellman2013prb}. In this method, both classical and quantum statistics for the phonon population can be conveniently considered. As in the case of MD simulations, isotope scattering is also considered. For details, see Ref. \cite{gu2019prb}.

\begin{figure}[htb]
\begin{center}
\includegraphics[width=\columnwidth]{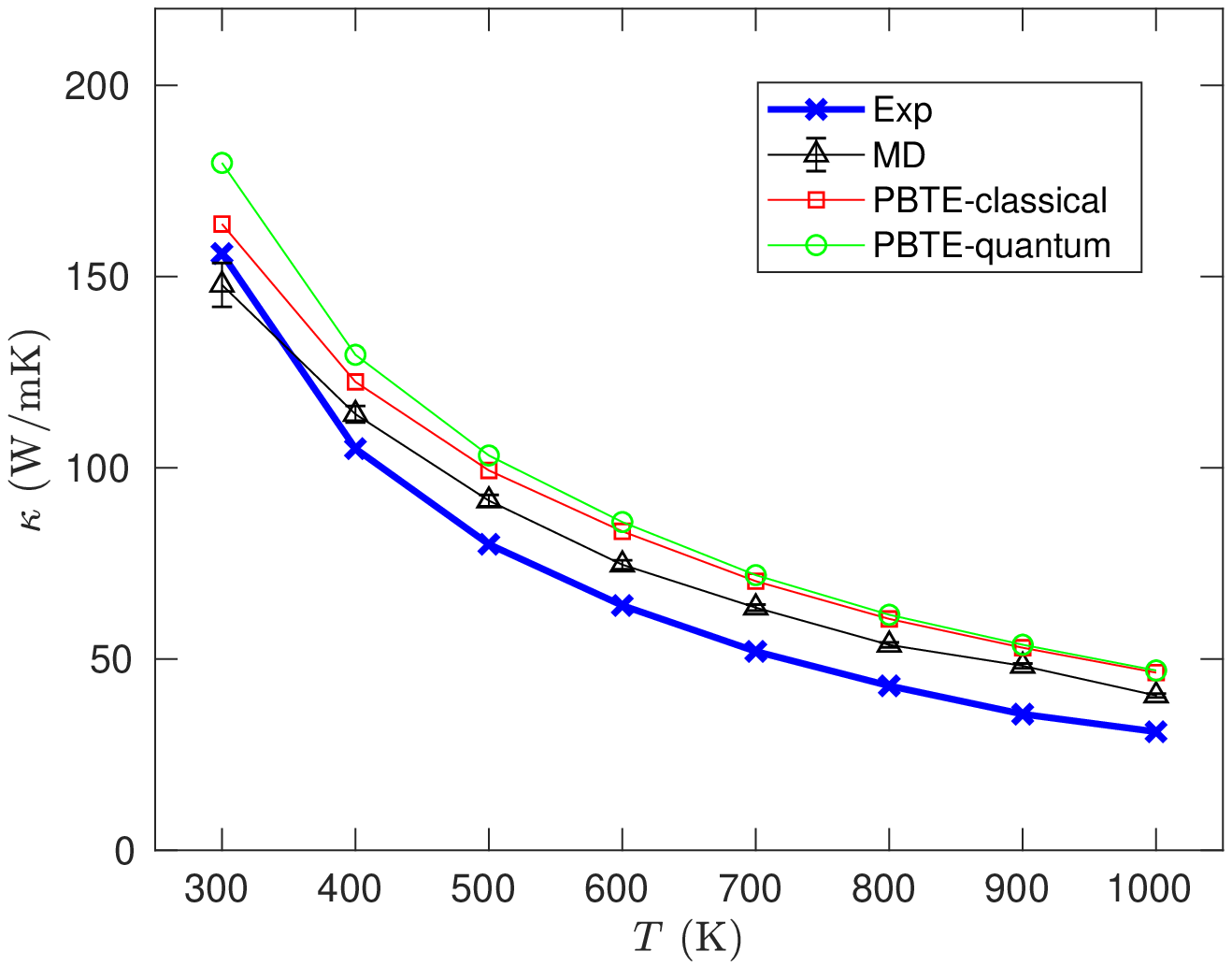}
\caption{Thermal conductivity as a function of temperature from the minimal Tersoff potential using MD simulation data and from PBTE calculations with classical and quantum statistics. Isotope scattering is included here. Experimental data are from Ref. \cite{Glassbrenner1964pr}. }
\label{figure:kappa_quantum}
\end{center}
\end{figure}

Figure \ref{figure:kappa_quantum} shows the classical and quantum thermal conductivity from the PBTE calculations using the minimal Tersoff potential, compared to the MD and experimental data. The thermal conductivity from PBTE calculations with classical statistics is slightly larger than that from MD, but they have a similar $T$ dependence. When quantum statistics is used in the PBTE calculations, the thermal conductivity at $T=300$ K increases by about $10 \%$. At temperatures above the Debye temperature (640 K), there is essentially no difference between the classical and quantum results. There are two competing quantum effects \cite{turney2009prb}: quantum statistics gives smaller modal heat capacities but larger phonon scattering times compared to classical statistics. In the temperature range considered here, the second effect is stronger, leading to underestimated $\kappa$ using classical statistics. Albeit, this overall effect is quite small (about $10 \%$) even at a temperature that is half of the Debye temperature. The point here is that if quantum corrections can be made to the classical MD results, the thermal conductivity at $T=300$ will be larger instead of smaller than the experimental value. Overall, we can conclude that our minimal Tersoff potential gives the best prediction for the thermal conductivity of diamond silicon among all the empirical potentials considered here.

\section{Summary and Conclusions\label{section:summary}}

In summary, we have proposed a minimal Tersoff empirical potential for diamond silicon and obtained a set of optimized parameters by fitting the potential against first-principles data using the genetic algorithm. The DFT data include energy and virial in many deformed structures and force in a few structures at finite temperatures. The optimized minimal Tersoff potential well describes the elastic constants, phonon dispersion, and thermal conductivity of diamond silicon simultaneously. Using classical statistics underestimates the thermal conductivity by an amount of about $10\%$ compared to using quantum statistics at room temperature. Both the fitting method and the optimized potential are made freely accessible from open-source codes we developed \cite{gpumd,gpuga}. The methods developed here are promising for constructing empirical potentials for new materials with good descriptions of the elastic and thermal properties.

\begin{acknowledgments}
ZF and TA-N acknowledge the supports from the National Science Foundations of China (NSFC) (No. 11974059) and from the Academy of Finland Centre of Excellence program QTF (Project 312298) and the computational resources provided by Aalto Science-IT project and Finland's IT Center for Science (CSC). YW, PQ and YS acknowledge the support from the financial support of National Key Research and Development Program of China (2016YFB0700500). XG acknowledges the support from the National Science Foundations of China (NSFC) (No. 51706134)
\end{acknowledgments}

\end{document}